\documentclass[%
 reprint,
 amsmath,amssymb,
 aps,
]{revtex4-1}

\usepackage{graphicx}
\usepackage{dcolumn}
\usepackage{bm}

\usepackage{float} 
\usepackage{url}
\usepackage{afterpage}

\begin{document}

\preprint{APS/123-QED}

\title{
  \large{Study of Tracking and Flavor Tagging\\ with FPCCD Vertex Detector} \vspace{0.3cm}
  \\  \scriptsize{"Talk presented at the International Workshop on Future \\Linear
  Colliders (LCWS13), Tokyo, Japan, 11-15 November 2013."}
}

\author{Tatsuya Mori$^1$}
 \email{moritatsu@epx.phys.tohoku.ac.jp}
\author{Daisuke Kamai$^1$, Akiya Miyamoto$^2$, Yasuhiro Sugimoto$^2$, Akimasa Ishikawa$^1$, Taikan Suehara$^3$, Eriko Kato$^1$, Hitoshi Yamamoto$^1$ }%
\affiliation{%
 Tohoku University$^1$,KEK$^2$,Kyushu University$^3$\\
}%





\begin{abstract}
One of the major physics goals at the ILC is the precise measurement
of the Higgs coupling constants to b-quarks and c-quarks.
To achieve this measurement, we need a high-performance
vertex detector leading to precise flavor tagging.
For this purpose, we are developing the Fine Pixel CCD (FPCCD) vertex detector.
In this paper, we will report on the development status of FPCCDTrackFinder, 
a new track finder improving tracking efficiency, especially in the low $p_T$ region,
and an evaluation result of the flavor tagging performance with 
FPCCDTrackFinder in the FPCCD vertex detector.  
\end{abstract}

\maketitle



\section{Introduction}

\subsection{Role of Vertex Detector in ILC}

\begin{figure}[!h]
	\centering
	\includegraphics[width=8cm]{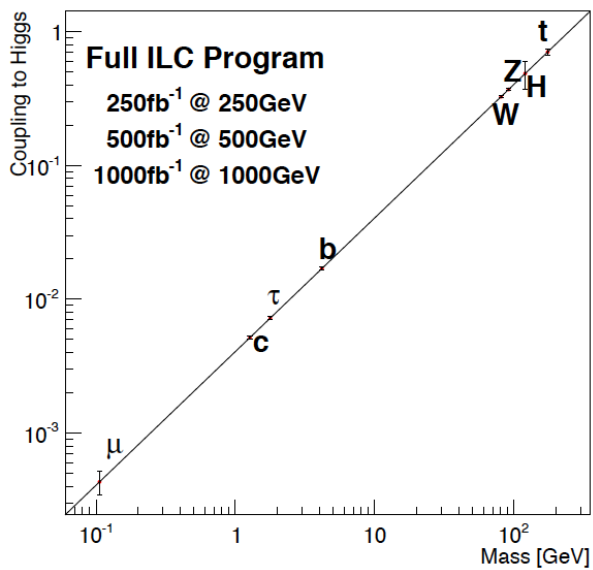}
	\caption[Expected measurement precision of Higgs coupling constant in ILC]
	{\small  Expected measurement precision of Higgs coupling constant in ILC.}
	\label{fig:Higgs_coupling}
\end{figure}

One of the major physics goals at the ILC is the precise measurement of
the Higgs coupling constants to b-quarks and c-quarks, which
is a critical test of Higgs and Yukawa interactions in the Standard Model (Figure \ref{fig:Higgs_coupling})~\cite{TDR:vol2}.
To achieve this measurement, we must identify Higgs decays into 
$b\Bar{b},\ c\Bar{c},\ or\ q\Bar{q}$ as precisely as possible, 
where q is a u-, d-, or s-quark. This identification is called flavor tagging.
One of the most discriminating variables for the flavor tagging is the difference 
between the secondary vertex position and the Higgs decay at the IP; 
typically, the proper decay length of b-hadron is $400 \sim 500\ \mathrm{\mu m}$, 
and that of c-hadron is $100 \sim 300\ \mathrm{\mu m}$.
Thus when the vertex resolution is improved, flavor tagging will be improved.
The vertex resolution is determined by the resolution of the impact parameter $d_0$ 
(in this study, we use the definition of track parameters in ~\cite{TrackParameter}),
which in the vertex detector at the ILC is required to satisfy 
\begin{equation}
   \sigma_{d_0} = 5 \mathrm{\mu m} \oplus \frac{10\mathrm{\mu m \cdot GeV/c}}{p \beta \sin^{3/2} \theta}  \label{IPReso_requirement}.
\end{equation}
In addition, the vertex detector is also required to keep a low occupancy of less than $2 \sim 3$\%. 
The dominant background in the vertex detector at the ILC is from $e^+e^-$ pairs generated from beamstrahlung.
The pair-BGs have relatively low momentum, so they curl many times through layers and increase pixel occupancy.     
To achieve these requirements and improve the flavor tagging, 
we are researching and developing the Fine Pixel CCD (FPCCD) Vertex Detector.
The FPCCD Vertex Detector is an optional detector for ILD detector concept~\cite{TDR:vol1},
and the performance was studied using the ILD software framework~\cite{ILCSoft}.

\subsection{FPCCD's Features}

FPCCD is a fully depleted silicon pixel sensor with very small pixel size~\cite{TDR:vol4}. As shown in Table \ref{table:fpccd_geometry},
the pixel size on layer-0 and layer-1 is $5\ \mathrm{\mu m}$, 
and that on the outer layers are $10\ \mathrm{\mu m}$.
\begin{table}
   \caption[FPCCD Geometry]{\small Geometry of the FPCCD VXD. }
   \begin{center}
      \begin{tabular}{c|r|r} \hline
         layer & \shortstack{distance \\from IP [mm]} & pixel size [$\mathrm{\mu m^2}$] \\ \hline \hline
         0     & 16                    & 5 $\times$ 5                    \\
         1     & 18                    & 5 $\times$ 5                    \\
         2     & 37                    & 10 $\times$ 10                  \\
         3     & 39                    & 10 $\times$ 10                  \\
         4     & 58                    & 10 $\times$ 10                  \\
         5     & 60                    & 10 $\times$ 10                  \\ \hline
      \end{tabular}
   \end{center}
   \label{table:fpccd_geometry}
\end{table}
The sensitive and total thickness of this sensor is 15 $\mathrm{\mu m}$ and 50 $\mathrm{\mu m}$, respectively.
The number of pixels of the FPCCD vertex detector is around 400 million. These features are expected to
lead to low occupancy and good vertex resolution.

\begin{figure}[!h]
	\centering
	\includegraphics[width=8cm]{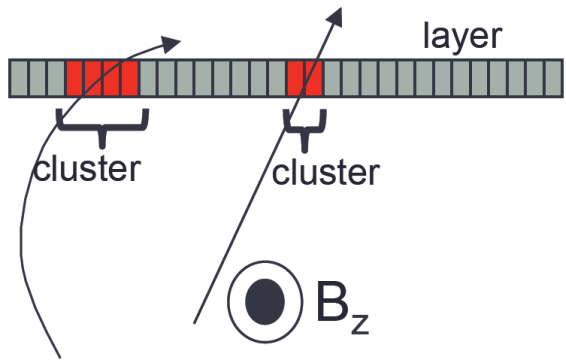}
	\caption[Explanation of Cluster]
	{\small  Explanation of Cluster. The red boxes denote pixels hit by particles. The pixels adjacent to each other are regarded as a cluster.}
	\label{fig:cluster_explanation}
\end{figure}

A notable feature of the very small pixels is that FPCCD can make clusters of hit pixels.
As shown in Figure \ref{fig:cluster_explanation}, when a particle goes through a layer,
it deposits energy into a few pixels in the layer. Those hit pixels can be regarded as a cluster.
The cluster shape depends on the way of traversing a layer, so the shape is useful for 
extrapolation in tracking and discrimination between pair-BG clusters and signal clusters.

Furthermore, by taking the weighted average of energy deposits of each pixel belonging to a cluster,
we can reconstruct the hit point more precisely.

The readout happens between trains, so the hit pixel data from all bunches in a train are accumulated in the FPCCD.
An advantage of this procedure is that we can ignore beam-induced RF noise, while it has the disadvantage that tracking
is challenging due to so many hit pixels.

\subsection{Occupancy and Impact Parameter Resolution}

We evaluated the pixel occupancy of the FPCCD vertex detector from pair-BGs using the ILC TDR beam parameters~\cite{TDR:vol1},
and the result is shown in Table \ref{table:occupancy}.
\begin{table}[!h]
   \caption[Pixel Occupancy]{
     \small Pixel occupancy of the FPCCD vertex detector on layer-0 from pair-BGs at the several center of mass energy.
     }
   \begin{center}
      \begin{tabular}{r|r|r|r} \hline
         $\mathrm{E_{CM}}$[GeV]& \multicolumn{1}{c}{\# of bunch crossings}& \multicolumn{2}{|c}{Pixel Occupancy} [\%]                  \\ \cline{3-4}
                               & \multicolumn{1}{c}{per train}            & \multicolumn{1}{|c}{layer-0} & \multicolumn{1}{|c}{layer-5}\\ \hline \hline
         250                   & 1312                                     & 0.561                        &0.009                        \\
         350                   & 1312                                     & 0.702                        &0.011                        \\
         500                   & 1312                                     & 1.244                        &0.020                        \\
         1000                  & 2450                                     &12.752                        &0.089                        \\ \hline
      \end{tabular}
   \end{center}
   \label{table:occupancy}
\end{table}
The occupancy on layer-0 for collision energies up to 500 GeV is relatively low, but at 1000 GeV it is too high.
The solutions for 1000 GeV are still being studied.

\begin{figure}[!h]
	\centering
	\includegraphics[width=8cm]{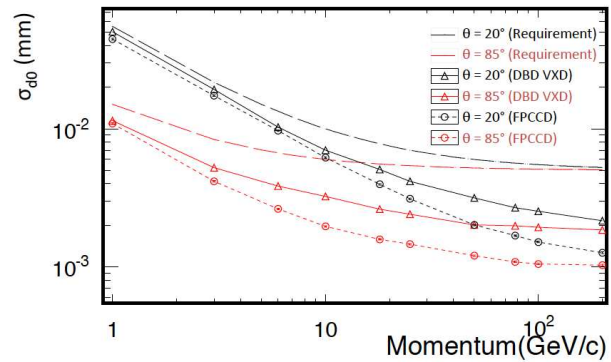}
	\caption[Impact Parameter Resolution]
	{\small Impact parameter resolution of the DBD ILD VXD (solid line) and the FPCCD VXD (dotted line) and the requirement of Equation \ref{IPReso_requirement} (long dashes).
  We assume the baseline position resolution given in Table \ref{table:position_resolution_DBD_FPCCD}.}
	\label{fig:IPReso_result}
\end{figure}

We also evaluated the impact parameter resolution of the FPCCD vertex detector. 
Since the FPCCD has very good position resolution, the impact parameter
resolution is expected to be good. The result of the study of the impact parameter resolution of the FPCCD is shown in
Figure \ref{fig:IPReso_result} ~\cite{TDR:vol4}.
For comparison, the configuration of the ILD VXD used in the Detailed Baseline Design (DBD)~\cite{DBD} is also evaluated. 
The position resolution of the DBD ILD VXD and the FPCCD VXD is shown in
Table \ref{table:position_resolution_DBD_FPCCD}.
\begin{table}
  \caption[Position Resolution of the DBD ILD VXD and the FPCCD VXD]{\small Position Resolution of the DBD ILD VXD and the FPCCD VXD in each layer.}
  \begin{center}
     \begin{tabular}{l|r|r} \hline
       layer &  \multicolumn{2}{|c}{Position Resolution [$\mathrm{\mu m}$]}   \\ \hline
             &  \multicolumn{1}{|c|}{DBD}    & \multicolumn{1}{c}{FPCCD}     \\ \hline 
       0     &  2.8                           & 1.4                           \\
       1     &  6.0                           & 1.4                           \\
       2     &  4.0                           & 2.8                           \\
       3     &  4.0                           & 2.8                           \\
       4     &  4.0                           & 2.8                           \\
       5     &  4.0                           & 2.8                           \\ \hline
     \end{tabular}
  \end{center}
  \label{table:position_resolution_DBD_FPCCD}
\end{table}
The FPCCD satisfies the requirement of Equation \ref{IPReso_requirement}, and gives around 1 $\mathrm{\mu m}$ $d_0$ resolution in the high momentum region.

\afterpage{\clearpage}

\section{FPCCDTrackFinder, A New Tracking Algorithm}

For studying the flavor tagging performance using FPCCD,
the tracking efficiency has been evaluated. 
The ILD tracking algorithm used in the DBD~\cite{DBD} was unable to efficiently reconstruct
low $p_T$ tracks (less than around 1.7 GeV/c) though there were enough VXD hits.
We have developed FPCCDTrackFinder, a new tracking algorithm
for reconstructing low $p_T$ tracks with high efficiency. In this section, firstly,
the DBD ILD tracking algorithm will be introduced briefly, and 
the result of evaluating the tracking efficiency will be shown.
Secondly, the difference between FPCCDTrackFinder and the DBD ILD tracking is explained.
Thirdly, the result of evaluating the tracking efficiency of 
FPCCDTrackFinder is shown. Finally, the amount CPU time and the memory consumption
are mentioned.

\subsection{The DBD ILD Tracking}\label{subsection:DBD_ILD_tracking}

\begin{figure}[!h]
	\begin{center}
    \centering
		\includegraphics[width=7cm]{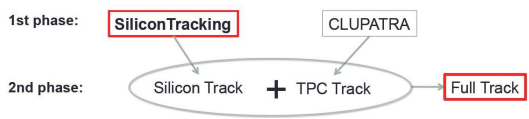}
	\end{center}
	\caption[Strategy of the DBD ILD tracking]
	{\small  Strategy of the DBD ILD tracking. After making a silicon track using only silicon tracker hits and a TPC track using only TPC hits,
  we combine a silicon track and a TPC track into a full track.}
	\label{fig:ILD_tracking}
\end{figure}

The strategy of the DBD ILD tracking is shown in Figure \ref{fig:ILD_tracking}.
In the first phase, the SiliconTracking processor reconstructs tracks using only VXD, SIT, and FTD hits,
and CLUPATRA processor reconstructs tracks using only TPC hits.
We refer to those tracks as "silicon track" and "TPC track". 
In the second phase, one silicon track and one TPC track are combined into one track, named full track. 
This full track is used for vertexing.

In this paper the tracking efficiency is defined by
\begin{gather}
   \mathrm{Tracking\ Efficiency} \equiv \notag \\ 
   \frac{\#\ \mathrm{of\ Good\ Track\ originated\ from\ Good\ Particle}}
   {\#\ \mathrm{of\ Good\ Particle}}  \label{definition_tracking_efficiency}
\end{gather}
where "Good Track" is a reconstructed track with VXD hits $\ge$ 5 and track purity $>$ 0.75,
and "Good Particle" is a Monte Carlo particle making VXD hits $\ge$ 6 and SIT hits $\ge$ 4.
Track purity is defined by 
\begin{equation}
   \mathrm{track\ purity}  \equiv 
   \frac{\#\ \mathrm{of\ true\ hits\ belong\ to\ the\ track}}
   {\#\ \mathrm{of\ hits\ belong\ to\ the\ track}}
\end{equation}
where true hits mean the hits originated from a particle corresponding to the track.

\begin{figure}[!h]
	\begin{center}
	  \includegraphics[width=7cm]{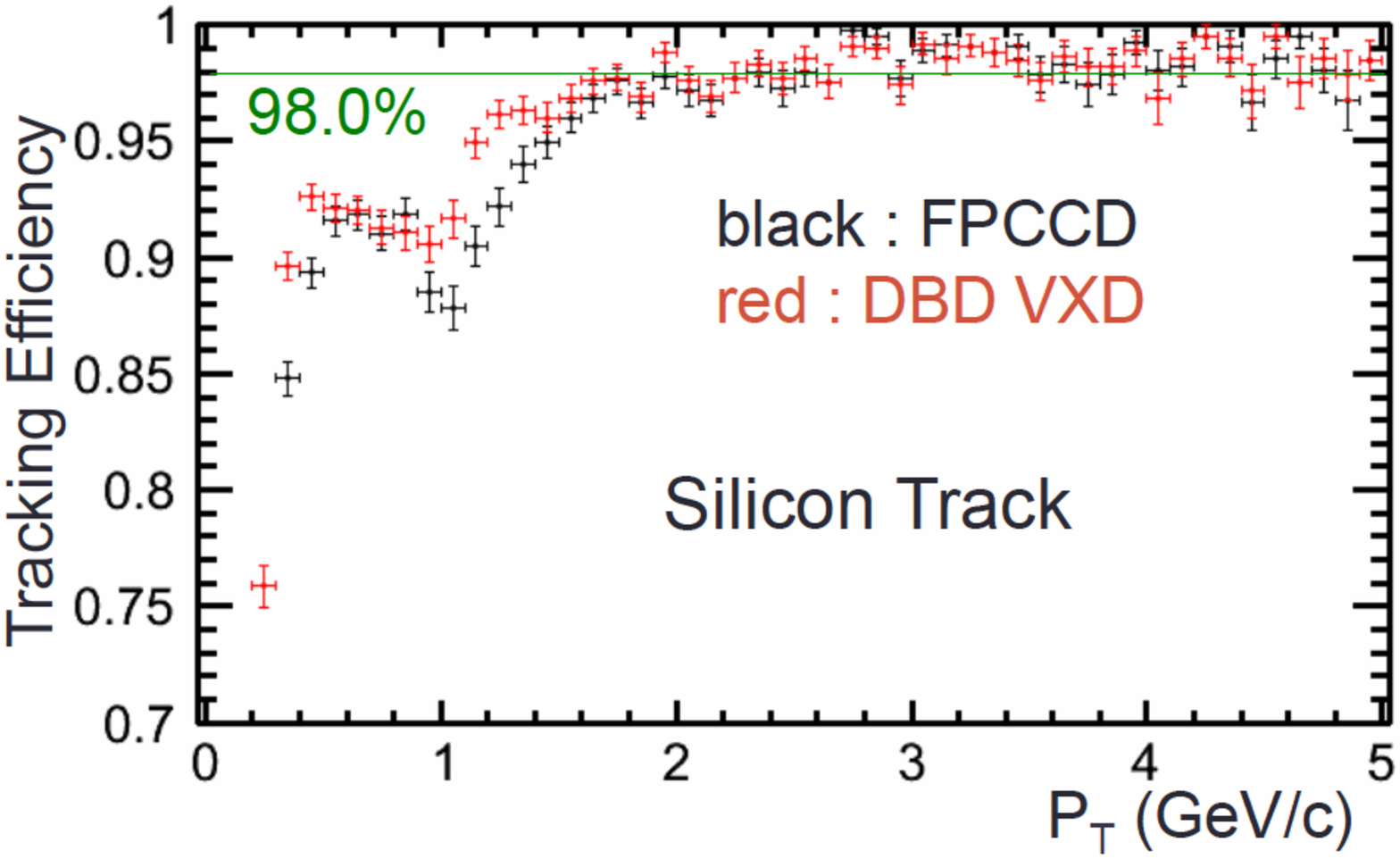}
	  \caption[Tracking Efficiency with DBD ILD tracking (silicon track)]
	  {\small  The tracking efficiency vs. $p_T$ with DBD ILD tracking (silicon track). }
	  \label{fig:trkeff_ildtrk_si}
	  \includegraphics[width=7cm]{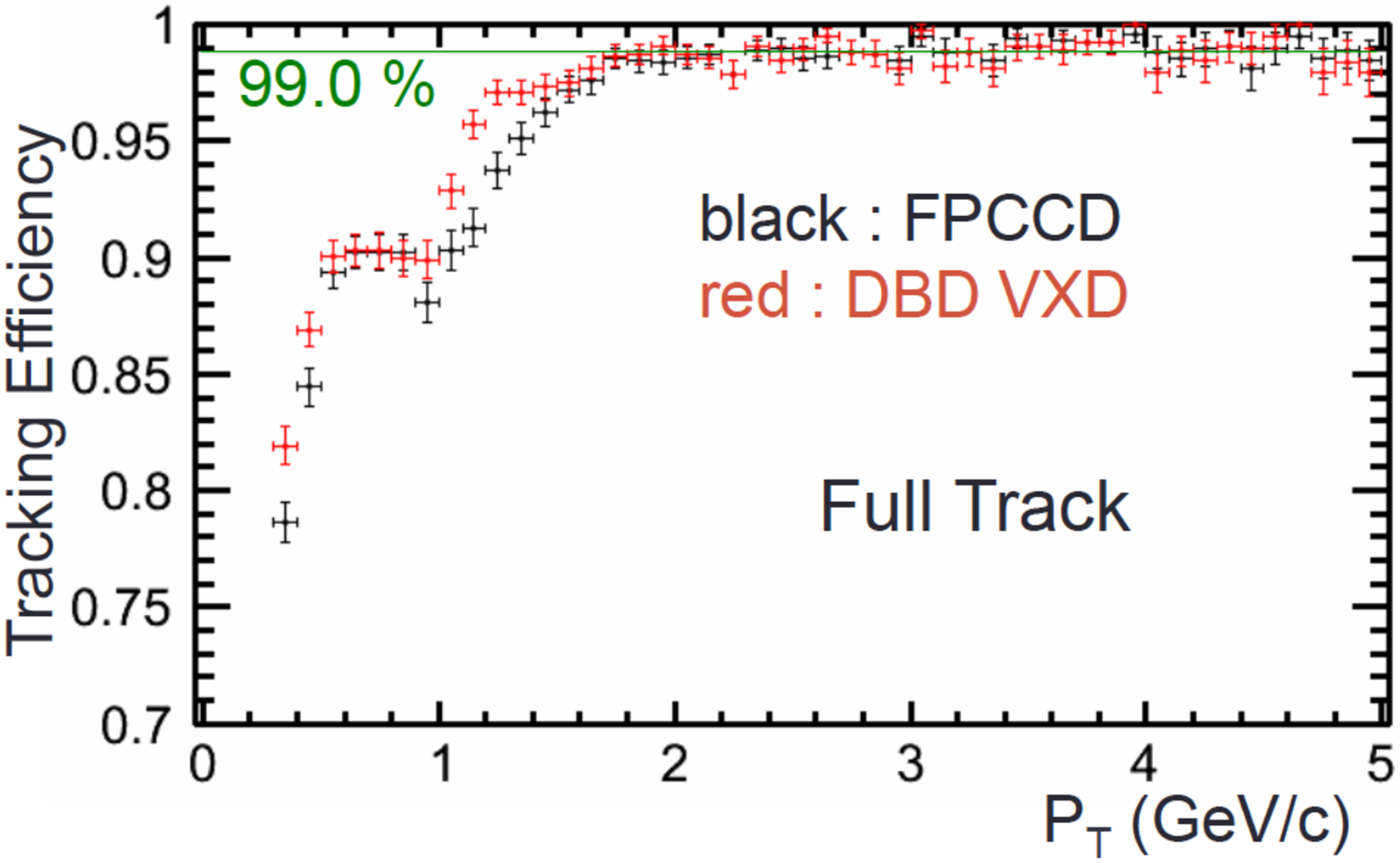}
	  \caption[Tracking Efficiency with DBD ILD tracking (full track)]
	  {\small  The tracking efficiency vs. $p_T$ with DBD ILD tracking (full track). }
	\end{center}
	\label{fig:trkeff_ildtrk_full}
\end{figure}

The sample used here is $t\Bar{t} \to 6jets$ at 350 GeV, 
and the number of events is 1000.
The results of evaluating the tracking efficiency with the DBD ILD tracking 
are shown in Figure \ref{fig:trkeff_ildtrk_si} (efficiency of silicon track) 
and Figure \ref{fig:trkeff_ildtrk_full} (efficiency of full track). 
As shown in Figure \ref{fig:trkeff_ildtrk_si} and Figure \ref{fig:trkeff_ildtrk_full}, 
the tracking efficiency of silicon and full tracks is around 98 and 99\% in $p_T$ $>$ 1.7 GeV/c, 
but decreases below $p_T$ = 1.7 GeV/c in both cases of the DBD ILD VXD and the FPCCD VXD. 
The reason of slight increase from 98\% (silicon track) to 99\% (full track) is 
that after combining a silicon and a TPC track into a full track, hits unused for tracking at this stage are
used to add further full tracks. However, in spite of this process, 
the efficiency of finding a full track deteriorates at the same $p_T$ as for silicon tracks.
This means that the efficiency of finding a silicon track contributes to 
the full track reconstruction efficiency, because the number of silicon tracks decreases at $p_T$ $<$ 1.7 GeV/c.
Thus, to improve tracking efficiency in the low $p_T$ region, we need to improve the silicon tracking.
For this purpose, FPCCDTrackFinder, a new tracking algorithm, has been developed.

\subsection{Differences between FPCCDTrackFinder and SiliconTracking of the DBD ILD Tracking}
FPCCDTrackFinder is based on the DBD ILD tracking, so we shall explain 
how FPCCDTrackFinder works by comparing SiliconTracking of the DBD ILD tracking.

\begin{figure}[!h]
	\begin{center}
		\includegraphics[width=7cm]{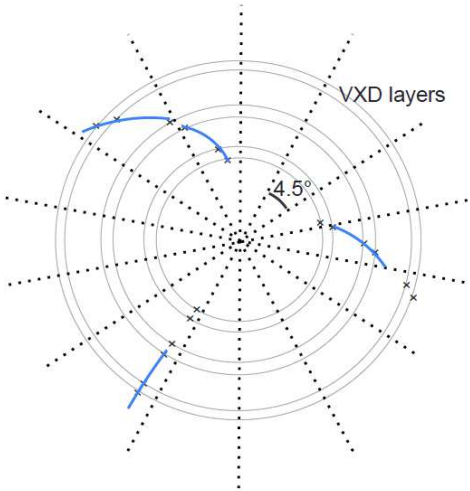}
	\end{center}
	\caption[Track Seeding Process]
	{\small  The graphical representation of the track seeding process. Crosses denote the VXD hits. 
  Blue curved lines denote the track seeds generated by this process.
  Track seeds are generated only in pieces divided by dotted line.}
	\label{fig:track_seed}
\end{figure}

Firstly, SiliconTracking generates track seeds (Figure \ref{fig:track_seed}). 
To simplify the graphical representation, the sectional view of only the VXD 
is shown in Figure \ref{fig:track_seed}, but actually the SIT is also used in SiliconTracking.
At the beginning, as shown in Figure \ref{fig:track_seed}, we divide the whole region into 80 pieces 
(namely $4.5^\circ$ per piece). Then, if there are 3 hits in each of 3 determined layers in a piece, 
simple helical fitting is applied and selected as a track seed, if the fit succeeds. 

Combinations of the 3 determined layers are as follows.
\begin{center}
(8\ 6\ 5)\ \ (8\ 6\ 4)\ \ (8\ 6\ 3)\ \ (8\ 6\ 2)\ \ (8\ 5\ 3)\ \ (8\ 5\ 2) 

(8\ 4\ 3)\ \ (8\ 4\ 2)\ \ (6\ 5\ 3)\ \ (6\ 5\ 2)\ \ (6\ 4\ 3)\ \ (6\ 4\ 2)

(6\ 3\ 1)\ \ (6\ 3\ 0)\ \ (6\ 2\ 1)\ \ (6\ 2\ 0)\ \ (5\ 3\ 1)\ \ (5\ 3\ 0)

(5\ 2\ 1)\ \ (5\ 2\ 0)\ \ (4\ 3\ 1)\ \ (4\ 3\ 0)\ \ (4\ 2\ 1)\ \ (4\ 2\ 0)
\end{center}
where 8 and 6 denote the outer and inner layer of the SIT, and numbers from 5 to 0
denote VXD layers. 

\begin{figure}[!h]
	\begin{center}
		\includegraphics[width=7cm]{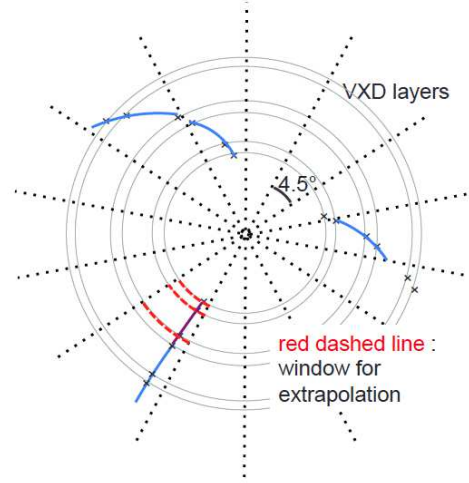}
	\end{center}
	\caption[Extrapolation Process]
	{\small  The graphical representation of the extrapolation process. 
  If there is a track seed in a piece ($4.5^\circ$), and if there
  are hits in the inner layers in the piece (namely, red dashed line), 
  the track seed is extrapolated with those hits. 
  }
	\label{fig:extrapolation}
\end{figure}

After the track seeding process, SiliconTracking extrapolates tracks 
from track seeds as shown in Figure \ref{fig:extrapolation}.
Like the track seeding process, the extrapolation process begins 
with the division of the whole region into 80 pieces. Next, if
at least one track seed exists in one of the 80 pieces, the 
following processes will be done with respect to the track seed.
\begin{enumerate}
\item[(1)] 
On a layer in the piece (namely, in the red line shown in Figure \ref{fig:extrapolation}),
we try to find the hit closest to the track seed.
\item[(2)]
If the distance between the hit and the track seed is less than a given threshold,
after adding the hit into the track seed, 
we apply simple helical fitting to the combined track seed
and the track is renewed if the fit succeeds.
\item[(3)] 
We redo (1) and (2) with respect to the other layers.
\end{enumerate}
That is the basic way that track seeds grow up to be well-reconstructed tracks.

However, there are two shortcomings in the track seeding process.
One is that the width of a piece, $4.5^\circ$, is too narrow
to generate a low $p_T$ track seed. In addition, the way of dividing the whole region leads to
tracks crossing the region boundary not being able to generate their track seeds.
The reason why widening pieces is not preferred is that the number of ghost track seeds and CPU time increase.
The other is that there are many combinations of the 3 layers.
This creates many ghost track seeds and consumes a lot of CPU time.

Furthermore, there are two known shortcomings of the extrapolation process.
One is that SiliconTracking uses simple helical fitting for extrapolation. 
This fitting doesn't consider multiple scattering and energy loss effect.
Low $p_T$ tracks tend to be affected relatively significantly by these effect, so
$\mathrm{\chi^2/ndf}$ from this fitting tends to become improperly high.
The third is that the search window for extrapolation is not flexible.
Since the search region does not depend on the processed track seed but on the piece width,
some hits for a low $p_T$ track seed cannot exist in the search region because
low $p_T$ tracks tend to cross the region boundary.

\begin{figure}[!h]
	\centering
	\includegraphics[width=7cm]{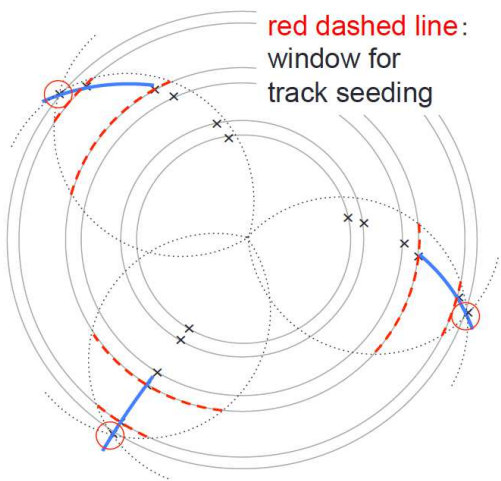}
	\caption[Track Seeding Process in FPCCDTrackFinder]
	{\small  The graphical representation of track seeding process in FPCCDTrackFinder. Crosses denote the VXD hits. 
  Crosses surrounded by red circles in the outer layer are used to determine a wide enough search region to
  catch track seeds with $p_T$ $>$ 0.18 GeV/c (by default).
  Dotted curved lines denote tracks passing through the IP with $p_T$ $=$ 0.18 GeV/c.
  Red dashed lines denote search windows for generating track seeds, and their length is determined by
  intersections between layers and the dotted curved lines.
  Blue curved lines denote the track seeds generated by this process.
  }
	\label{fig:track_seed_modified}
\end{figure}

For overcoming these shortcomings, FPCCDTrackFinder generates track seeds as 
shown in Figure \ref{fig:track_seed_modified}.
Firstly, FPCCDTrackFinder chooses one of the hits on the outer layer among given 3 layers for creating track seeds.
Secondly, FPCCDTrackFinder creates all candidate track seeds whose hits on the middle and inner layers exist in a
search window calculated in the following way.
\begin{itemize}
\item[(1)]
We draw on the plane perpendicular to the uniform magnetic field two hypothetical tracks generated at the IP, going through the hit on the outer layer, 
and having $p_T$ = 0.18 GeV/c (by default) and +1 and -1 electric charge respectively.
\item[(2)]
The two points of intersection between the middle layer among given 3 layers for creating track seeds and the two tracks are regarded as
the end points of the search window. The same goes for the inner layer.
\end{itemize}
As compared to SiliconTracking, FPCCDTrackFinder hardly fails to reconstruct low $p_T$ track seeds due to the crossings of the region boundary.

In addition, the combinations of 3 layers are reduced from the ones for DBD ILD tracking to 
\begin{center}
(8\ 6\ 5)\ \  (8\ 6\ 4)\ \  (8\ 5\ 4)\ \  (6\ 5\ 4)\ \  (5\ 4\ 3)\ 
\end{center}
This reduction of combinations is appropriate because the above search window 
can catch almost all track seeds with $p_T$ $>$ 0.18 GeV/c 
Notice that the inner layers of the VXD, 0 to 2-layer, are not used in order not to increase 
the number of ghost track seeds and the CPU time, because the hit density of these layers is relatively high.

\begin{figure}[!h]
	\centering
	\includegraphics[width=7cm]{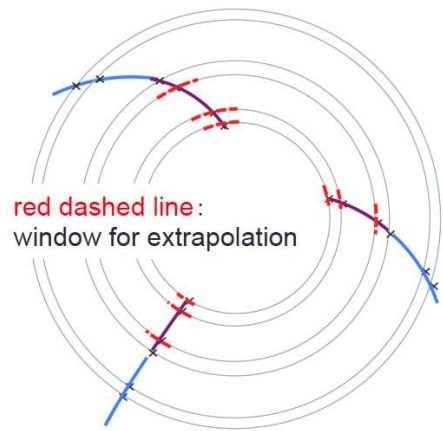}
	\caption[Extrapolation process in FPCCDTrackFinder]
	{\small  The graphical representation of the extrapolation process in FPCCDTrackFinder. 
  FPCCDTrackFinder determines the width of the search window (namely, red dashed line) 
  by considering track parameters and the errors of the processed track seed.  
  }
	\label{fig:extrapolation_FPCCDTF}
\end{figure}

After track seeding process, FPCCDTrackFinder extrapolates tracks from track seeds as 
shown in Figure \ref{fig:extrapolation_FPCCDTF}.
Instead of simple helical fitting, FPCCDTrackFinder uses a Kalman Filter, a fitter
considering multiple scattering and energy loss effects~\cite{KalmanFilter}. Thus, although the computational budget 
increases, a Kalman Filter can output $\chi^2/ndf$ more accurately than simple helical fitting for
low $p_T$ tracks.

In addition, FPCCDTrackFinder does not divide 
the whole region into 80 pieces, but instead determines the width 
of the search window by considering track parameters and the errors
of the processed track seed (Figure \ref{fig:extrapolation_FPCCDTF}). 
This method can catch true hits and properly reduce extra time for considering false hits.
If we use FPCCD VXD and FPCCDTrackFinder, due to the good position resolution, 
the search window can be narrowed further and the probability of mis-extrapolation decreases.

\subsection{Tracking Efficiency of FPCCDTrackFinder}
\begin{figure}[!h]
  \begin{center}
	  \includegraphics[width=7cm]{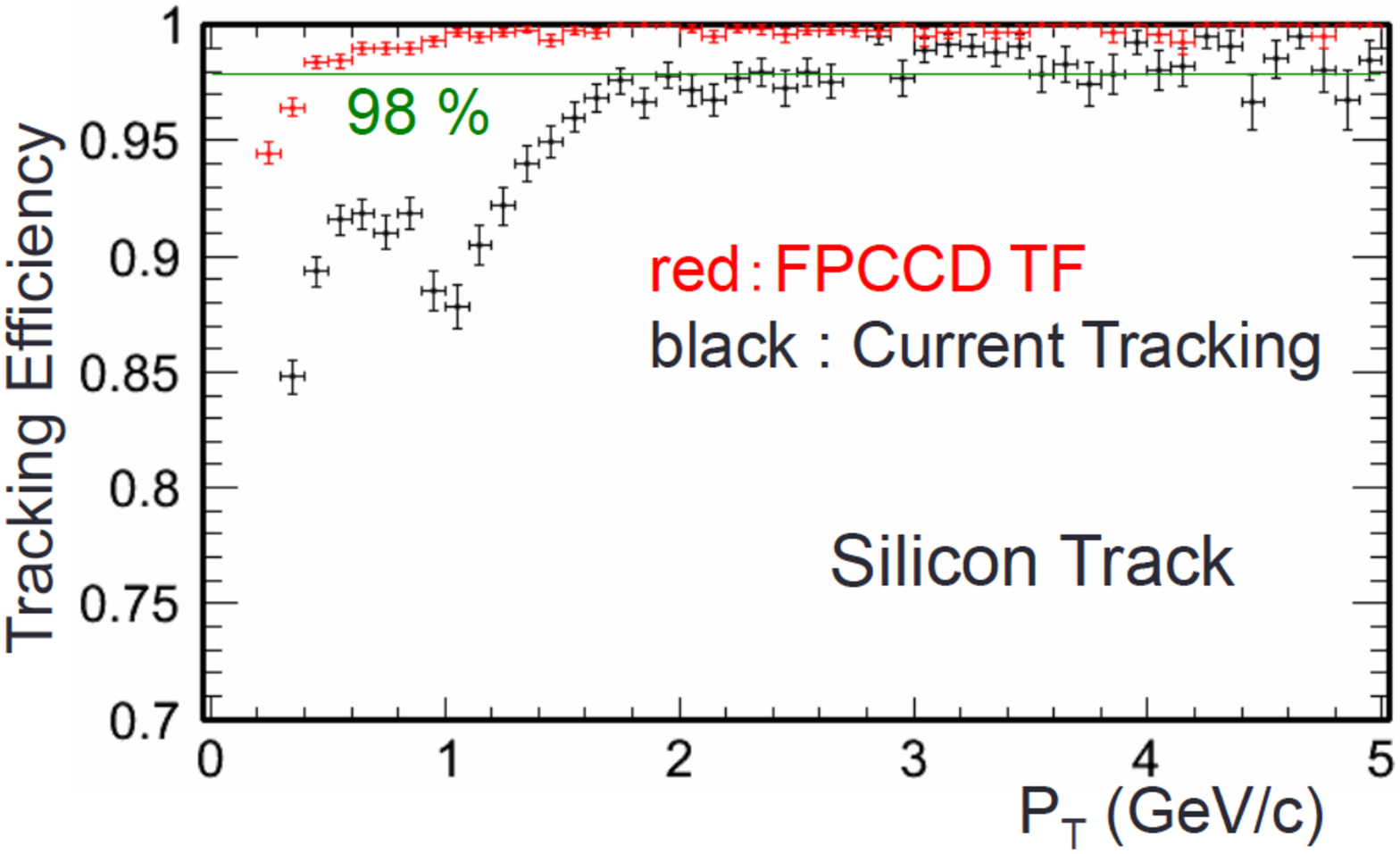} 
	  \caption[FPCCDTrackFinder VS DBD ILD tracking with FPCCD (silicon track, $p_T$)]
	  {\small  The tracking efficiency vs. $p_T$ of silicon tracks with the FPCCD VXD and the DBD ILD tracking (black crosses) and FPCCDTrackFinder (red crosses).}
	  \label{fig:trkeff_ildtrk_vs_fpccdtf_si}
	  \includegraphics[width=7cm]{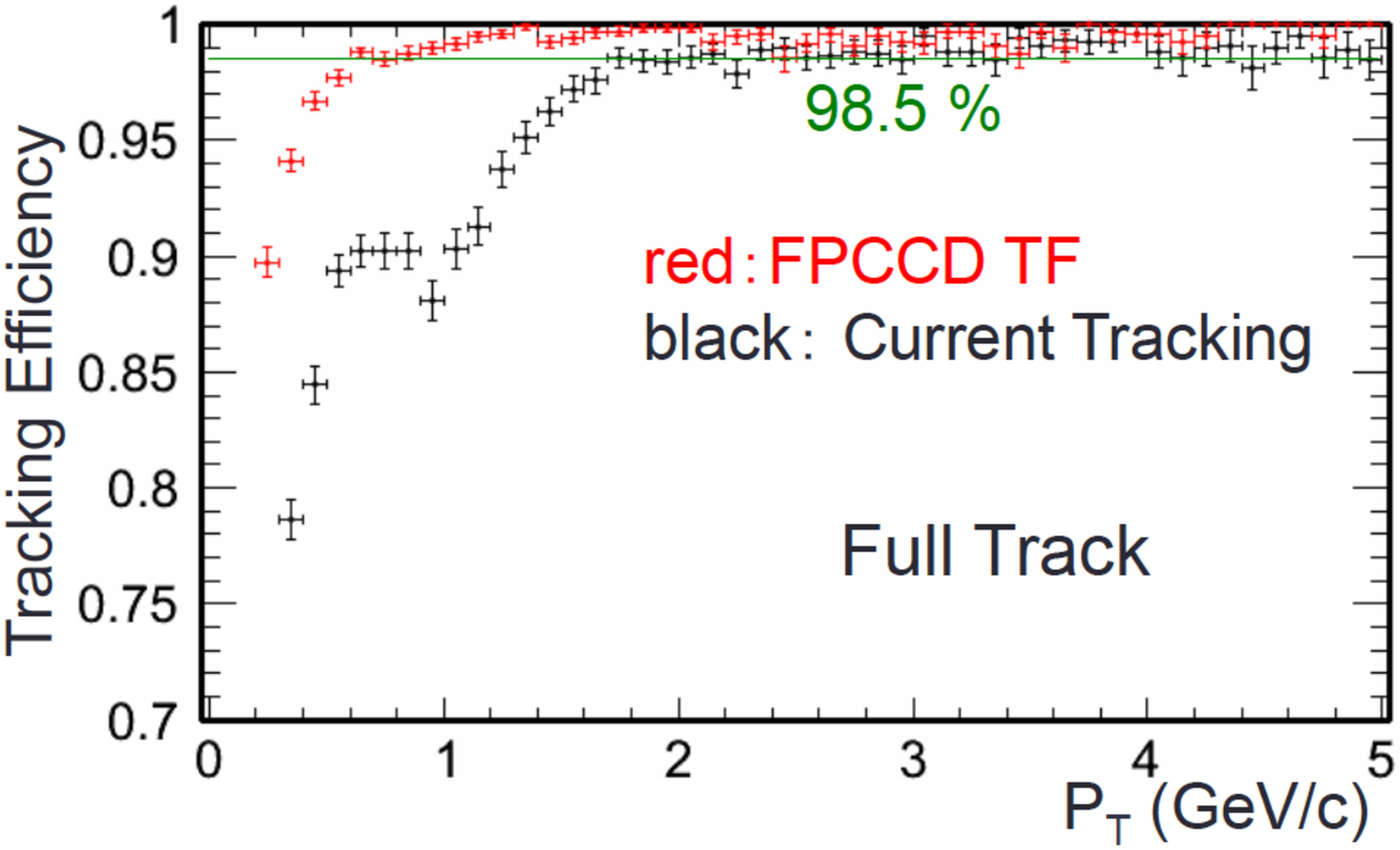} 
	  \caption[FPCCDTrackFinder VS DBD ILD tracking with FPCCD (full tracks, $p_T$)]
	  {\small  The tracking efficiency vs. $p_T$ of full tracks with the FPCCD VXD and the DBD ILD tracking (black crosses) and FPCCDTrackFinder (red crosses). }
	  \label{fig:trkeff_ildtrk_vs_fpccdtf_full}
  \end{center}
	\centering
\end{figure}

The setup for the evaluation of the tracking efficiency 
is the same as described in section \ref{subsection:DBD_ILD_tracking}.
Comparisons of the DBD ILD tracking and FPCCDTrackFinder with FPCCD VXD are
shown in Figure \ref{fig:trkeff_ildtrk_vs_fpccdtf_si} (efficiency of reconstructing silicon tracks) and 
Figure \ref{fig:trkeff_ildtrk_vs_fpccdtf_full} (efficiency of reconstructing full tracks).
\begin{figure}[!h]
	\includegraphics[width=7cm]{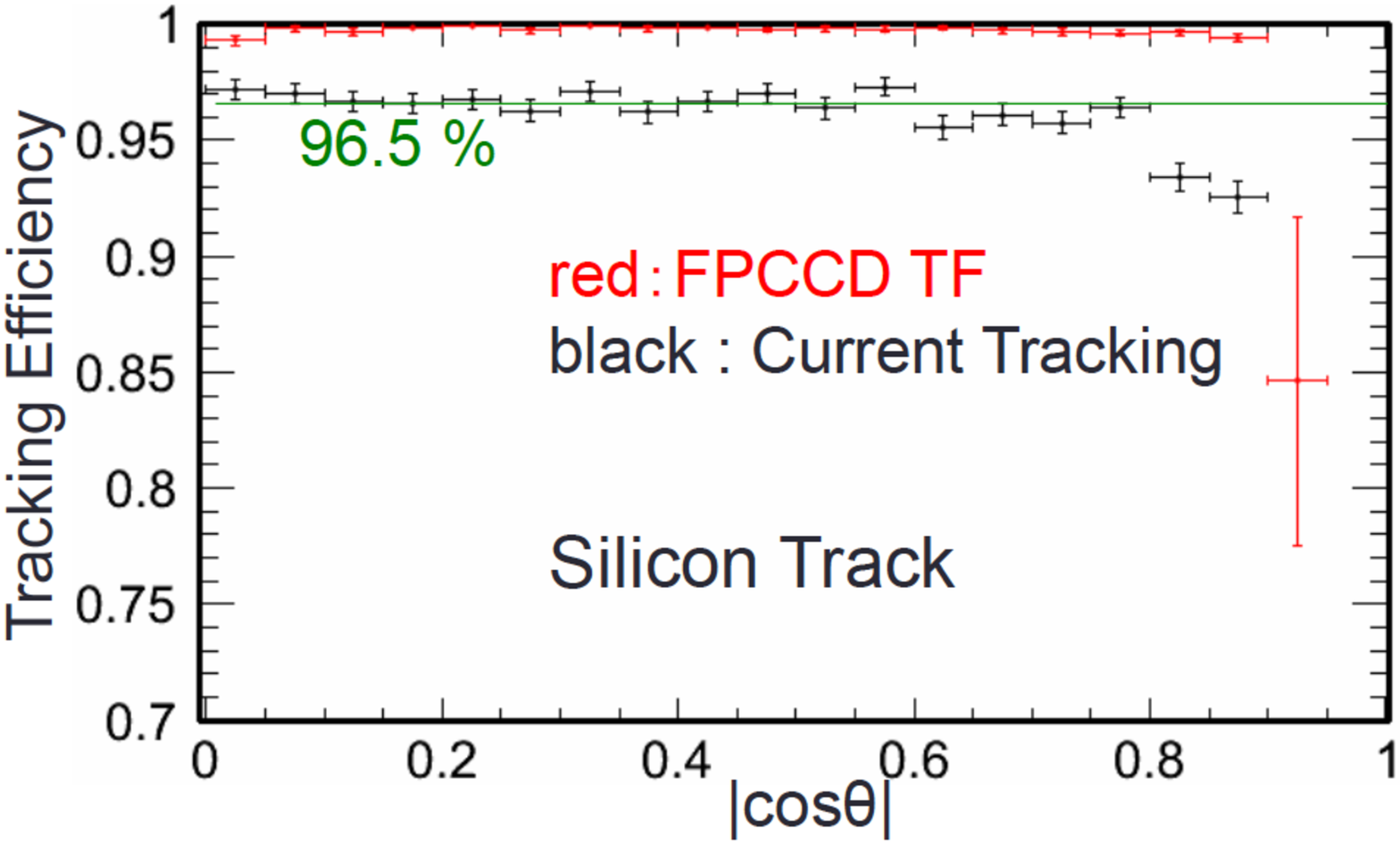} 
	\caption[FPCCDTrackFinder VS DBD ILD tracking with FPCCD (silicon track, $\cos\theta$)]
	{\small  The tracking efficiency vs. $\cos\theta$ of silicon tracks ($|p|$ $>$ 1 GeV/c) 
  with the FPCCD VXD and the DBD ILD tracking (black crosses) and FPCCDTrackFinder (red crosses).
  }
	\label{fig:trkeff_ildtrk_vs_fpccdtf_si_cos}
	\centering
	\includegraphics[width=7cm]{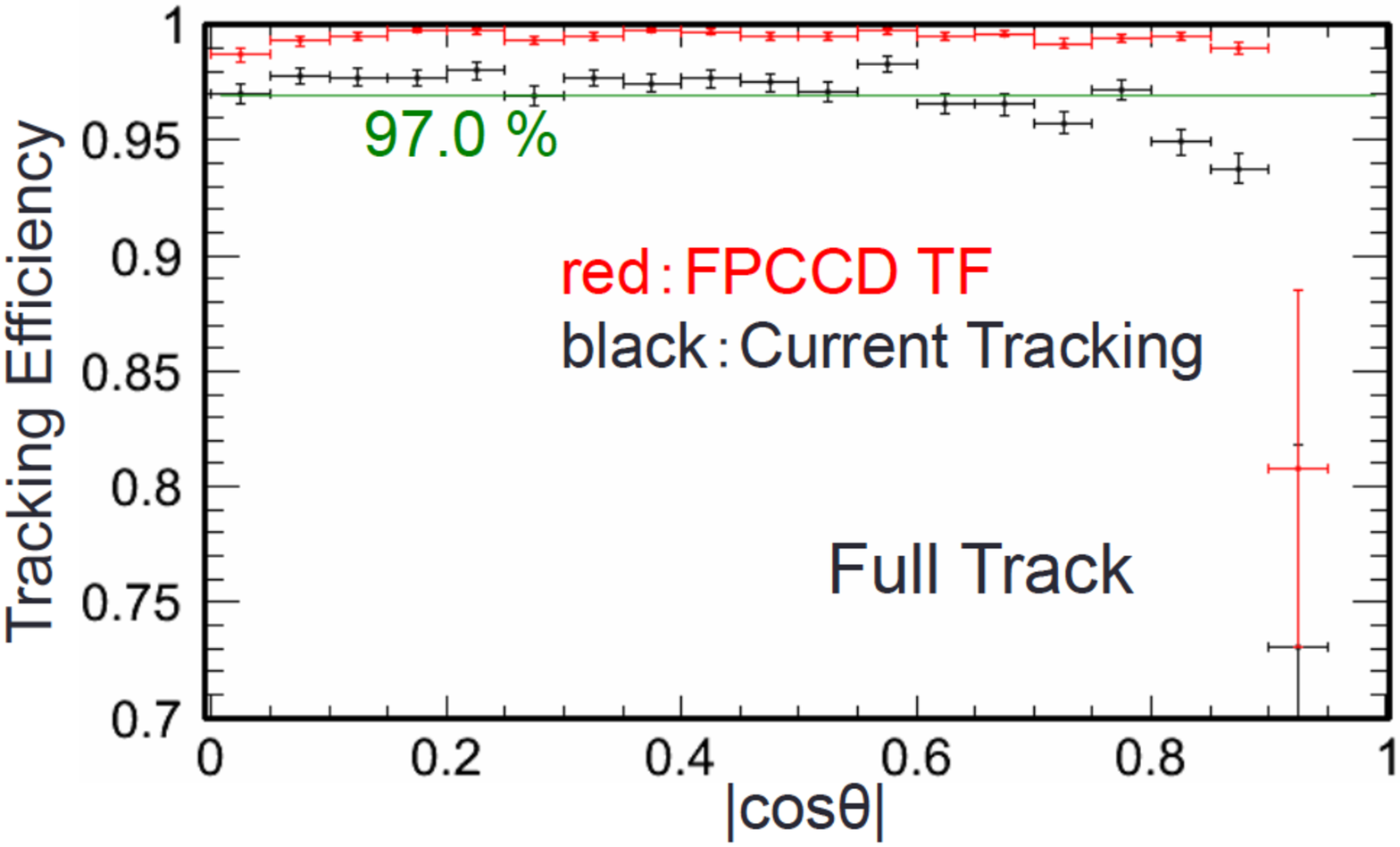} 
	\caption[FPCCDTrackFinder VS DBD ILD tracking with FPCCD (full tracks, $\cos\theta$)]
	{\small  The tracking efficiency vs. $\cos\theta$ of full tracks ($|p|$ $>$ 1 GeV/c)
  with the FPCCD VXD and the DBD ILD tracking (black crosses) and FPCCDTrackFinder (red crosses). 
  }
	\label{fig:trkeff_ildtrk_vs_fpccdtf_full_cos}
\end{figure}
As shown in Figure \ref{fig:trkeff_ildtrk_vs_fpccdtf_si} and Figure \ref{fig:trkeff_ildtrk_vs_fpccdtf_full},
FPCCDTrackFinder improves the efficiency of both silicon and full tracks
to $\sim$ 99\% above $p_T$ = 0.6 GeV/c, as compared to 97\% with the DBD ILD tracking.
In addition, dependency of the efficiency on $\cos\theta$ is shown in 
Figure \ref{fig:trkeff_ildtrk_vs_fpccdtf_si_cos} (silicon tracks)
and Figure \ref{fig:trkeff_ildtrk_vs_fpccdtf_full_cos} (full tracks),
where tracks with $|p|$ $>$ 1 GeV/c are evaluated.
These Figures show that FPCCDTrackFinder improves the tracking efficiency of both silicon and full track 
to $\sim$ 99\% within $|\cos\theta|$ = 0.9, as compared to 96.5\% with the DBD ILD tracking.
The reason that the efficiency decreases in $|\cos\theta| > 0.9$ is that
the acceptance of the SIT is $|\cos\theta| < 0.9$, so the efficiency in $|\cos\theta| > 0.9$ 
is not considered in this evaluation.

\begin{figure}[!h]
	\centering
	\includegraphics[width=7cm]{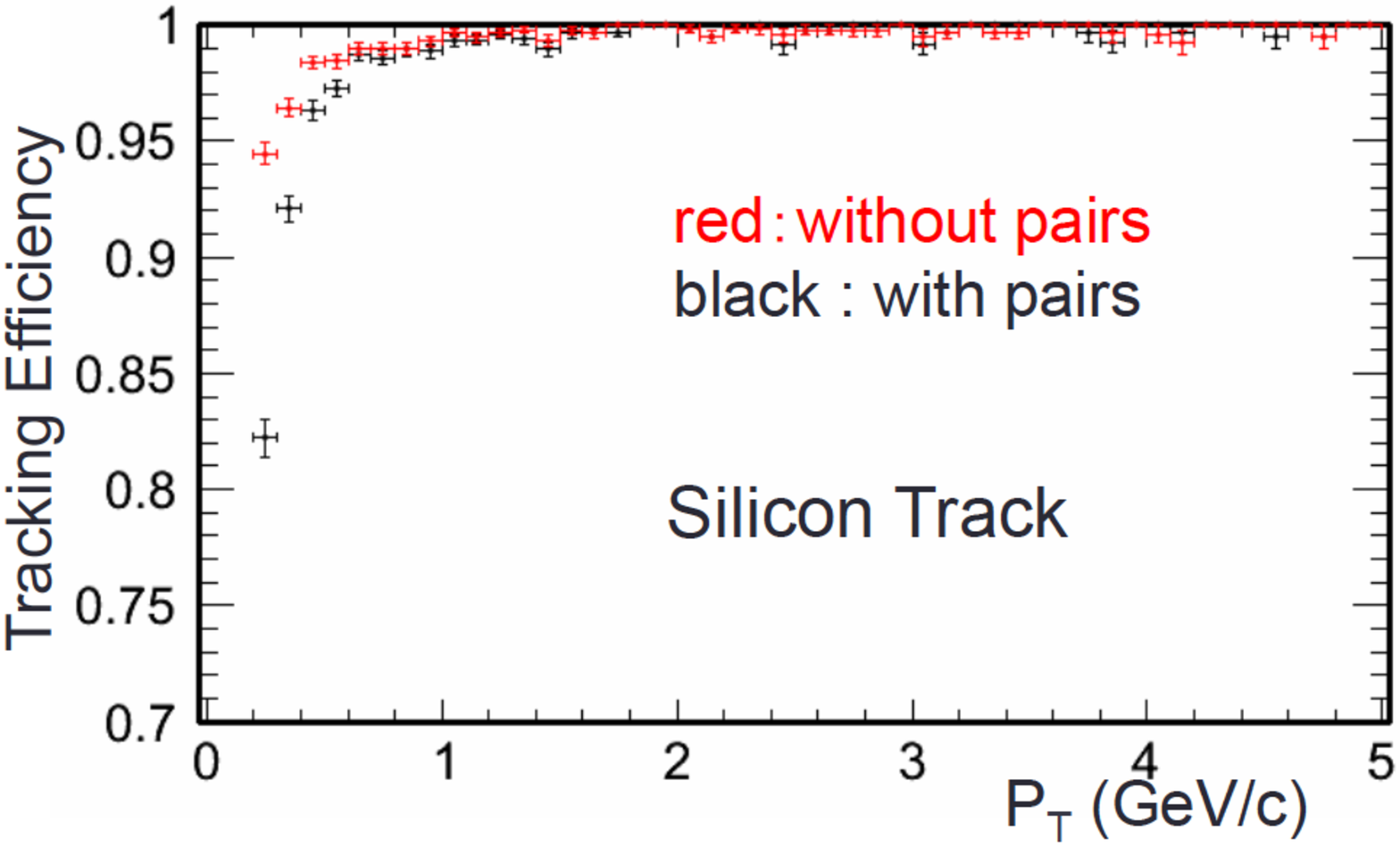} 
	\caption[FPCCDTrackFinder and FPCCD without and with pair-BG (silicon track, $p_T$)]
	{\small  The tracking efficiency vs. $p_T$ of silicon tracks with the FPCCD VXD and FPCCDTrackFinder with pair-BG (black crosses) and without pair-BG (red crosses).}
	\label{fig:trkeff_fpccdtf_with_pairs_si}
	\centering
	\includegraphics[width=7cm]{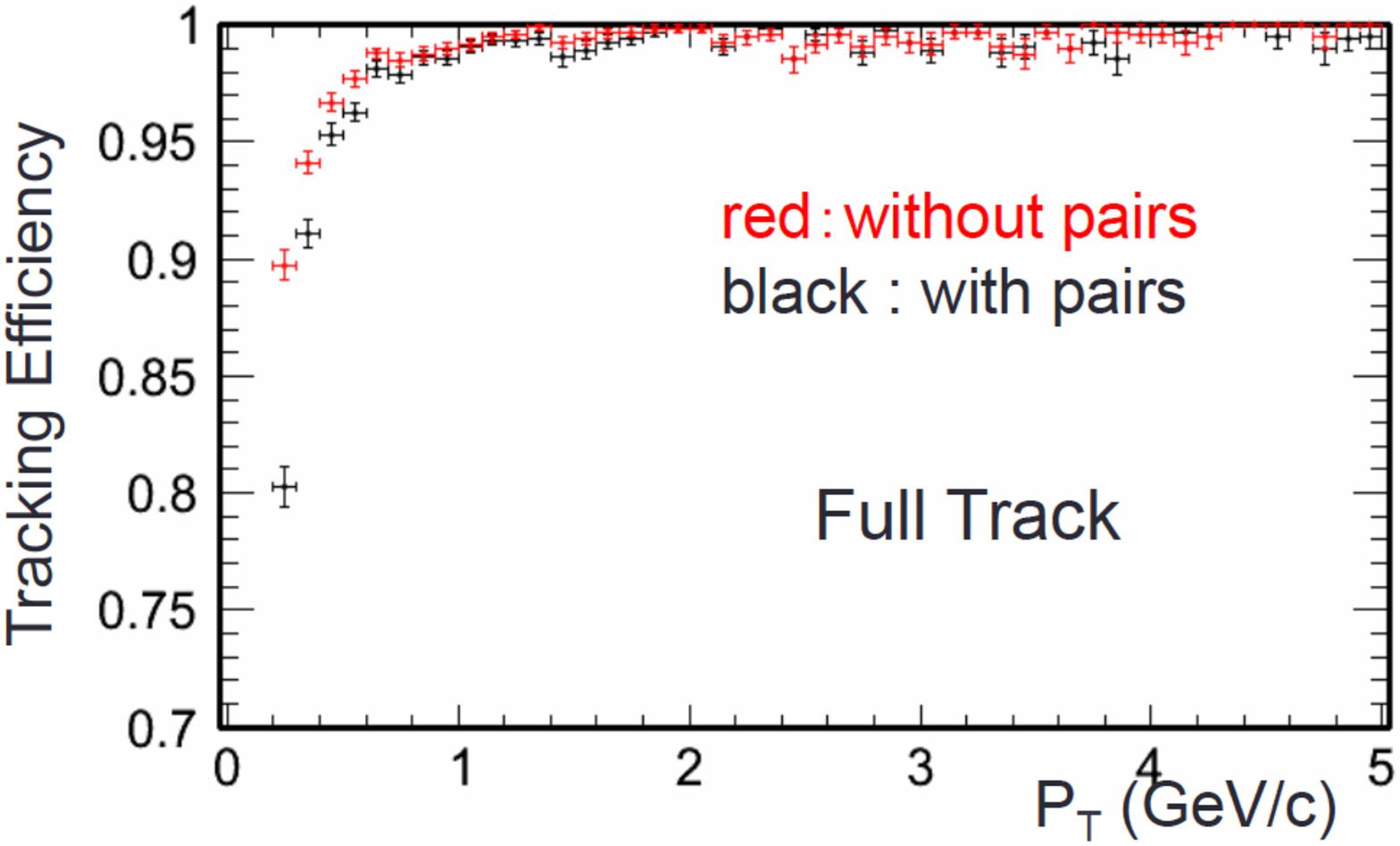} 
	\caption[FPCCDTrackFinder and FPCCD without and with pair-BG (full track, $p_T$)]
	{\small  The tracking efficiency vs. $p_T$ of full tracks with the FPCCD VXD and FPCCDTrackFinder with pair-BG (black crosses) and without pair-BG (red crosses).}
	\label{fig:trkeff_fpccdtf_with_pairs_full}
\end{figure}

Next, the result of the evaluation of the tracking efficiency with pair-BG at 350 GeV is shown in
Figure \ref{fig:trkeff_fpccdtf_with_pairs_si} (silicon track) 
and Figure \ref{fig:trkeff_fpccdtf_with_pairs_full} (full track).
\begin{figure}[!h]
	\centering
	\includegraphics[width=7cm]{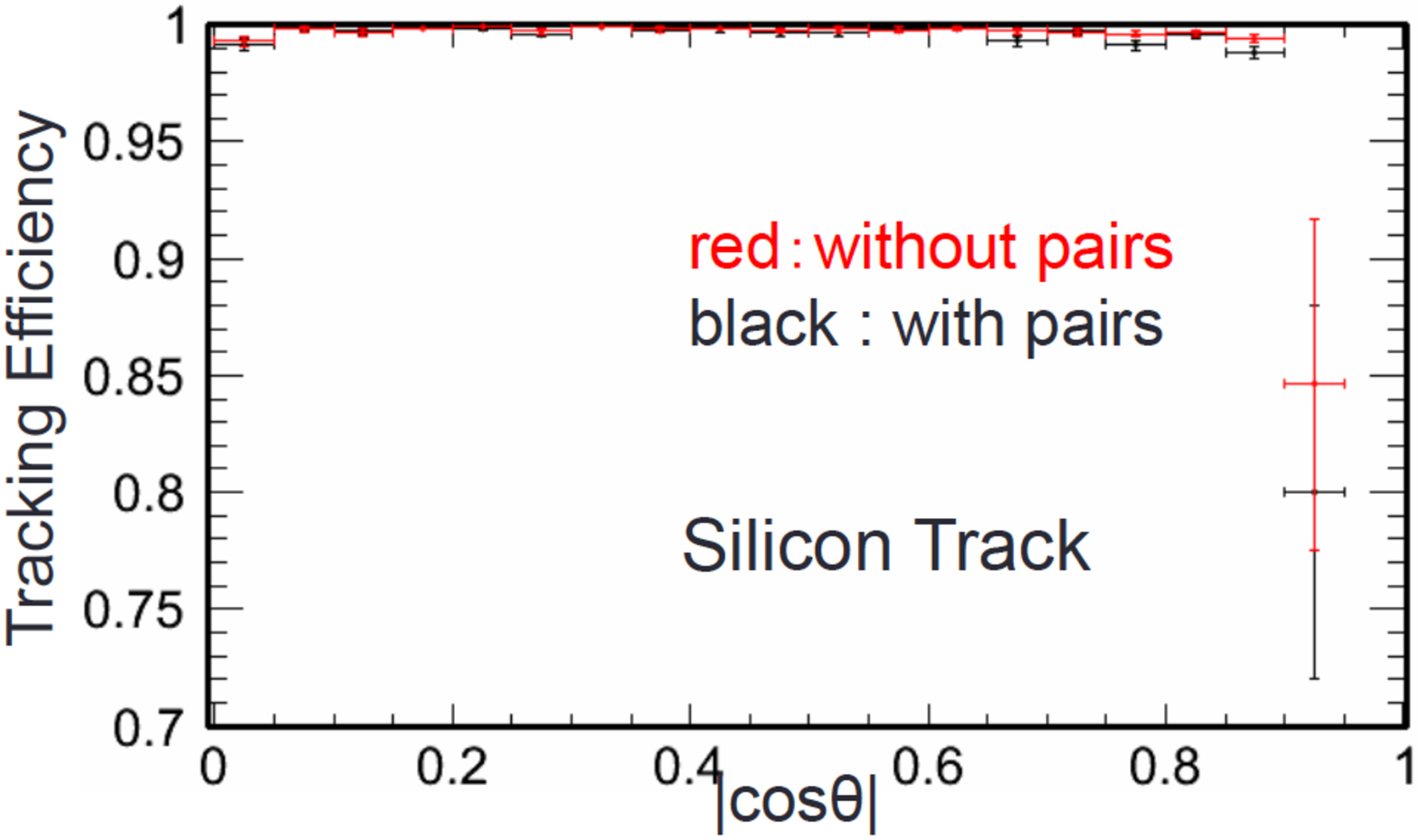} 
	\caption[FPCCDTrackFinder and FPCCD without and with pair-BG (silicon track, $\cos\theta$)]
	{\small  The tracking efficiency vs. $\cos\theta$ of silicon tracks ($|p|$ $>$ 1 GeV/c)
  with the FPCCD VXD and FPCCDTrackFinder with pair-BG (black crosses) and without pair-BG (red crosses).
  }
	\label{fig:trkeff_fpccdtf_with_pairs_si_cos}
	\centering
	\includegraphics[width=7cm]{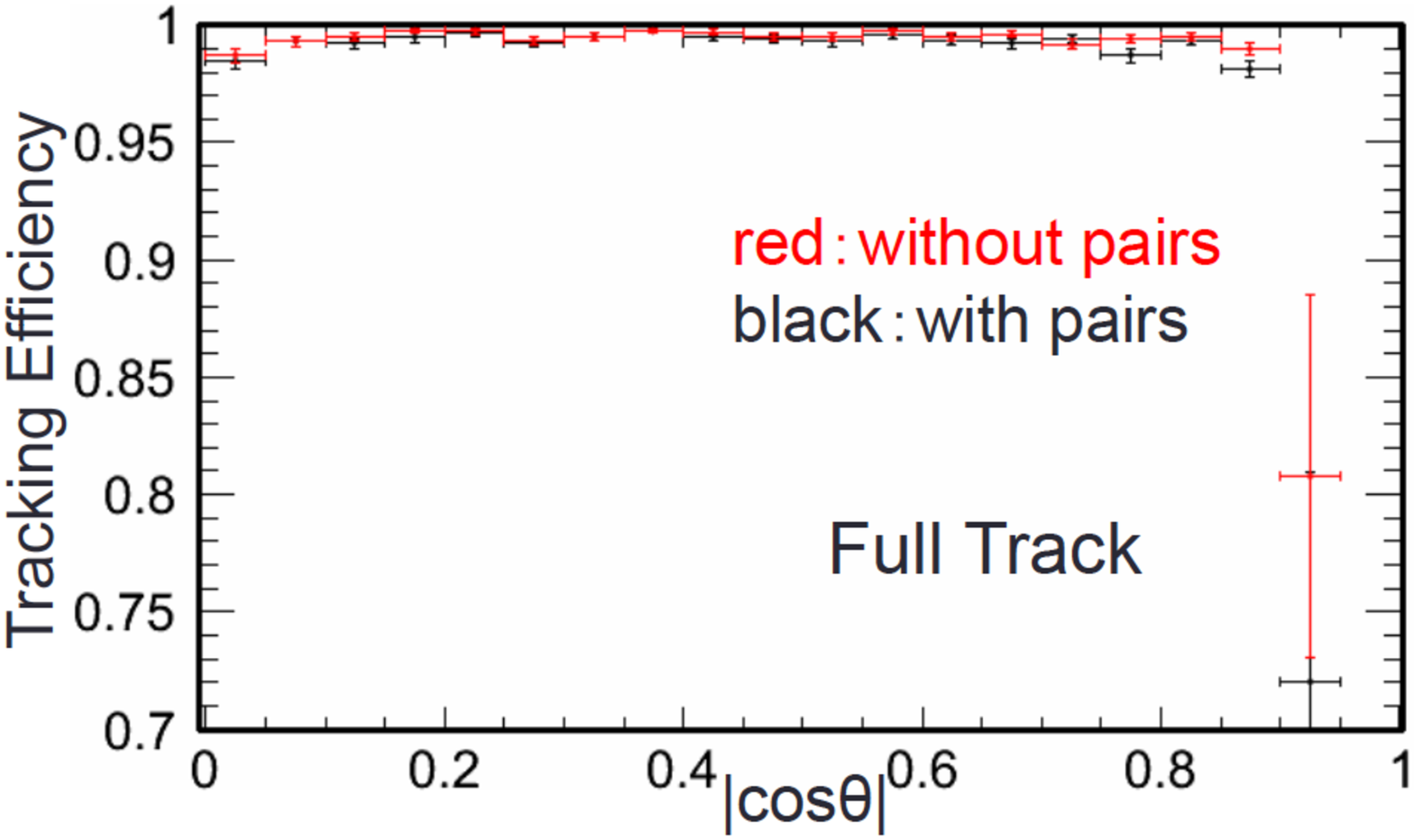} 
	\caption[FPCCDTrackFinder and FPCCD without and with pair-BG (full track, $\cos\theta$)]
	{\small  The tracking efficiency vs. $\cos\theta$ of full tracks ($|p|$ $>$ 1 GeV/c)
  with the FPCCD VXD and FPCCDTrackFinder with pair-BG (black crosses) and without pair-BG (red crosses).
  }
	\label{fig:trkeff_fpccdtf_with_pairs_full_cos}
\end{figure}
As shown in Figure \ref{fig:trkeff_fpccdtf_with_pairs_si} and 
Figure \ref{fig:trkeff_fpccdtf_with_pairs_full},
the tracking efficiency holds $\sim$ 99\% above $p_T$ = 0.6 GeV/c
regardless of the presence of pair-BG. 
In addition, the dependency of the efficiency on $\cos\theta$ 
in the presence of pair-BG is shown in 
Figure \ref{fig:trkeff_fpccdtf_with_pairs_si_cos} (silicon track)
and Figure \ref{fig:trkeff_fpccdtf_with_pairs_full_cos} (full track),
where tracks with $|p|$ $>$ 1 GeV/c are evaluated.
As shown in Figure \ref{fig:trkeff_fpccdtf_with_pairs_si_cos} and
Figure \ref{fig:trkeff_fpccdtf_with_pairs_full_cos},
the tracking efficiency holds $\sim$ 99\% within $|\cos\theta| = 0.9$ 
regardless of the presence of pair-BG.

\subsection{CPU Time and Memory}
It was difficult to reconstruct tracks using the FPCCD VXD in the presence of pair-BG
before FPCCDTrackFinder became available. This is due to the fact that the presence of pair-BG
increases the CPU time too much especially during the track seeding.
When $t\Bar{t} \to 6 jets$ with pair-BG at 350 GeV are reconstructed with FPCCDTrackFinder and the FPCCD
VXD, the CPU time for the tracking is around 3 hours per event, and the memory consumption peaks around 3.5 GB.
The CPU time of track seeding process occupies around 5/6 of the total CPU time of FPCCDTrackFinder.
Therefore, one of the strategies for further reduction of the CPU time is to improve the track seeding process.

\section{Performance Evaluation of Flavor Tagging}\label{section:FlavorTag}
In this section, the result of the evaluation of the flavor tagging performance is shown.
\subsection{Setup for the Evaluation}
The flavor tagging process is implemented by LCFIPlus~\cite{LCFIPlus}.
The sample for the evaluation is $Z \to b\Bar{b},\ c\Bar{c},\ q\Bar{q}\ @\ 91.2\ \mathrm{GeV}$,
with 25000 events for the testing, and 25000 events for the training.
LCFIPlus gives us the information of b-jet, c-jet, and q-jet 
likelihood for the reconstructed jets. 
The b-tag efficiency and the b-tag purity are defined by 
\begin{equation}
  \textrm{b-tag efficiency} \equiv \frac{\textrm{\#\ of\ b-tagged\ b-jet}}{\textrm{\#\ of\ all\ b-jet}}
\end{equation}
\begin{equation}
  \textrm{b-tag purity} \equiv \frac{\textrm{\#\ of\ b-tagged\ b-jet}}{\textrm{\#\ of\ all\ b-tagged\ jets}}
\end{equation}
The same goes for the c-tag and the q-tag.
The value of the purity depends on the branching ratio of $Z \to b\Bar{b},\ c\Bar{c},\ q\Bar{q}$, 
so in this paper, we assume~\cite{ParticleDataGroup}
\begin{align}
   BR(Z \to b\Bar{b}) &= 0.1512 \\
   BR(Z \to c\Bar{c}) &= 0.1203 \\
   BR(Z \to q\Bar{q}) &= 0.428
\end{align}
In this evaluation, we do not include pair-BG.

\subsection{Flavor Tagging with FPCCDTrackFinder}
\begin{figure}[H]
	\centering
	\includegraphics[width=8cm]{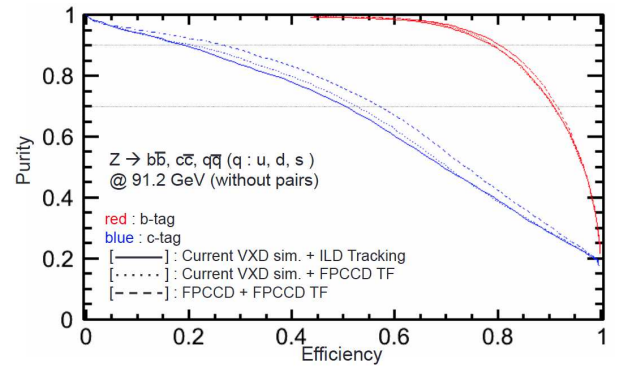}
	\caption[Flavor Tagging Performance]
	{\small  The evaluation of flavor tagging performance. Red line and blue line denote b-tag and c-tag performance respectively.
  The solid line shows the performance with DBD ILD VXD and the DBD ILD tracking,
  the dotted line using DBD ILD VXD and FPCCDTrackFinder, and
  the dashed line using FPCCD VXD and FPCCDTrackFinder. 
  }
	\label{fig:flavor_tag_zpole_result}
\end{figure}

In Figure \ref{fig:flavor_tag_zpole_result}, the red line shows the b-tag performance,
and the blue line shows the c-tag performance.
The solid line shows the performance with the DBD ILD VXD and the DBD ILD tracking,
the dotted line using the DBD ILD VXD and FPCCDTrackFinder, and
the dashed line using the FPCCD VXD and FPCCDTrackFinder. 
By comparing the solid line with the dotted line, we can see that
FPCCDTrackFinder improves the c-tag performance (for example, the c-tag efficiency improves by 2.5\%
in c-tag purity 70\%). Furthermore, by comparing the dotted line with the dashed line,
we can see that FPCCD VXD improves b-tag and c-tag performance (for example, b-tag efficiency
improves by 2\% in b-tag purity 90\%, and the c-tag efficiency improves by 4\% in c-tag purity 70\%). 
As a result, FPCCDTrackFinder and the FPCCD VXD improve the flavor tagging performance 
without considering pair-BG.

\section{Summary}

We have presented FPCCDTrackFinder, a new tracking algorithm.
FPCCDTrackFinder improves the tracking efficiency to $\sim$ 99\% 
in $p_T > 0.6$ GeV/c and $|\cos\theta| < 0.9$. 
FPCCDTrackFinder has enabled the tracking with FPCCD VXD in the presence of pair-BG
for the first time, and achieves a tracking efficiency of $\sim$ 99\%
in $p_T > 0.6$ GeV/c and $|\cos\theta| < 0.9$ regardless of the presence of pair-BG. 
We have also evaluated the performance  of flavor tagging in some cases.
FPCCDTrackFinder improves flavor tagging performance; c-tag efficiency
increases by 2.5\% for a c-tag purity of 70\%.
The FPCCD VXD also improves flavor tagging performance;
b-tag efficiency increases by 2\% for a b-tag purity of 90\%, and
the c-tag efficiency increases by 4\% for a c-tag purity of 70\%.

\section{Acknowledgments}

The authors would like to thank all the members of the ILC physics subgroup and the ILD group for useful
discussions, Dr. Jan Fridolf Strube for the detailed checks for this paper, 
and the JSPS grant-in-aid for specially promoted research No. 23000002.

\end{document}